\documentclass{appolb}
\usepackage{graphicx}

\begin{document}
\title{Recent results on charmonium-like (exotic) XYZ states at the BESIII/BEPCII experiment in Beijing/China 
\thanks{Presented at the $10^{\rm th}$ international workshop Excited QCD 2018}%
}
\author{F. Nerling\footnote{Email: F.Nerling@gsi.de}, \\on behalf of the BESIII Collaboration
  \address{Institut f\"ur Kernphysik, Goethe Universit\"at Frankfurt, \\ and  GSI Darmstadt, Germany}
}
\maketitle
\begin{abstract}
With about 12 fb$^{-1}$ collected XYZ data sets, BESIII continues the exploration of the
exotic charmonium-like states. In this talk, recent results on the measurements of the 
spin-parity determination of $Z_{\rm c}(3900)$, as well as on line-shapes of 
$e^+e^- \rightarrow J/\psi\,\pi\pi, h_{\rm c}\pi\pi, \psi(2S)\,\pi^0\pi^0/\pi^+\pi^-$, 
and $\pi^+ D^0 D^{*-}$ from open charm are discussed. Also, the recent observation of 
$e^+ e^- \rightarrow \phi \chi_{c1/2}$ at $\sqrt(s)=4.6$\,GeV is reported.
\end{abstract}
  
\section{Introduction}
In the charmonium region, the $c\bar{c}$ charmonium states can successfully be described using potential models.
All the predicted states have been observed with the expected properties beneath the open-charm threshold and 
excellent agreement is achieved between theory and experiment. 
Above the open-charm threshold, however, there are still many predicted states that have not yet been discovered, 
and, surprisingly, quite some unexpected states have been observed since 2003. Interesting examples of these so-called 
(exotic) charmonium-like ``XYZ'' states are the $X(3872)$ observed by Belle~\cite{X3872_belle}, the vector states 
$Y(4260)$ and $Y(4360)$, both discovered by BaBar using initial state radiation (ISR)~\cite{Y4260_barbar,Y4360_barbar}, 
and the charged state $Z_{\rm c}(3900)^\pm$ discovered by BESIII~\cite{Zc3900_besiii}, shortly after confirmed by 
Belle~\cite{Zc3900_belle}, that is a manifestly exotic state; for a recent overview see e.g.~\cite{reviewMitchel_etal_2016}. 

The Beijing Spectrometer (BES) at the Beijing Electron-positron Collider (BEPC) in China started initially in 1989, and the 
BESIII/BEPCII experiment~\cite{besiii} is the latest incarnation that began operation in March 2008 after major upgrades 
were finalised.

The multi-purpose detector allows for coverage of a broad hadron physics programme, including not only charmonium and open-charm 
spectroscopy but also electromagnetic form factor as well as $R$ scan measurements and many others.
We have collected the world largest data sets in the $\tau$-charm mass region. Among those are unique high-luminosity data sets 
to explore the still unexplained $XYZ$ states of in total more than 5\,fb$^{-1}$ accumulated above 3.8\,GeV.

\section{Recent major results on charmonium-like exotic XYZ states}
With BESIII/BEPCII conventional as well as charmonium-like (exotic) $XYZ$ states can be studied. 
In the $e^+e^-$ annihilation, we have direct access to vector $Y$ states ($J^{PC}$=$1^{--}$) that are 
produced at unprecedented statistics. Also, we can study charged as well as neutral $Z_{\rm c}$ states 
indirectly produced (together with recoil particles), whereas $X$ states are accessible via radiative 
decays, see {\it e.g.}~\cite{nerlingMorionds2017}. 

\subsection{Spin-parity determination of the $Z_{\rm c}(3900)$ and $Z_{\rm c}(3885)$}
\begin{figure}[tp!]
\vspace{-0.6cm}
    \begin{center}
     \includegraphics[clip, trim= 75 248 15 70,width=1.0\linewidth]{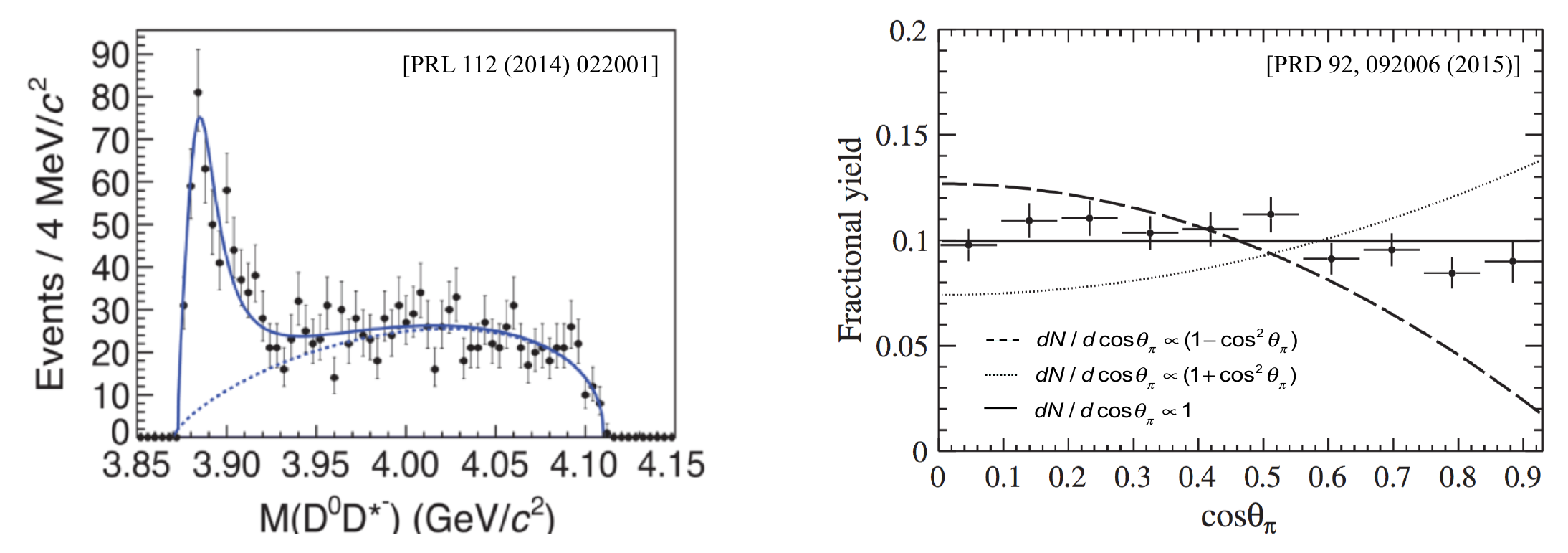}
    \end{center}
\vspace{-0.6cm}
      \caption{Results on the charged $Z_{\rm c}$ state observed in open charm decays.
        {\it Left:} Observation of $Z_{\rm c}(3885)^\pm\rightarrow D\bar{D^*}^\pm$ at $\sqrt{s}$=$4.26$\,GeV (525 pb$^{-1}$).
        {\it Right:} Angular distribution $|\cos\theta_{\pi}|$ in comparison to different $J^P$ assumptions based on data sets at 
        $\sqrt{s}$=$4.23, 4.26$\,GeV (1.9 fb$^{-1}$).
}
       \label{ZcSpinParity_a} 
\end{figure}
The discovery of the $Z_{\rm c}(3900)^\pm$ state is due to the charge in combination with the mass 
a strong hint for the first four-quark state being observed. 
After observation of the neutral partner $Z_{\rm c}(3900)^0\rightarrow J/\psi\pi^0$~\cite{Zc3900neutral}, confirming earlier evidence 
reported by CLEO-c~\cite{cleo-c_ZcNeutral}, a $Z_{\rm c}(3900)^{\pm,0}$ isospin triplet seems to be established. 
Furthermore, also a second isospin triplet $Z_{\rm c}(4020)^{\pm,0}$ has meanwhile been established in the BESIII 
data~\cite{Zc4020,Zc4020_neutral_BESIII}, also consistent with others~\cite{Zc4020_1_CLEO,Zc4020_b_Belle}. 
Despite this remarkable progress, the nature of these states is still unclear and the question is, whether the different 
decays the $Z_{\rm c}$ states have been observed in (hidden versus open charm) are decay modes of the same state. 
\begin{figure}[tp!]
\vspace{-0.6cm}
    \begin{center}
     \includegraphics[clip, trim= 75 28 15 290,width=1.0\linewidth]{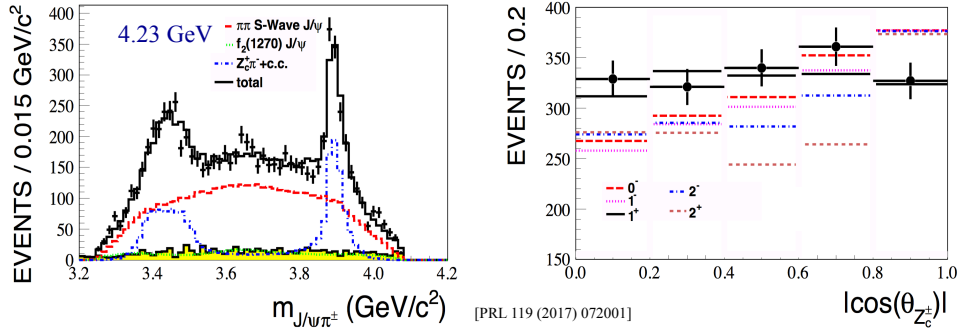}
    \end{center}
\vspace{-0.6cm}
      \caption{Results on the charged $Z_{\rm c}$ states observed in hidden charm decays.
        {\it Left:} Amplitude analysis result for $Z_{\rm c}(3900)^\pm\rightarrow J/\psi\pi^\pm$ based on data sets at 
        $\sqrt{s}=4.23$\,GeV (similar at 4.26\,GeV).
        {\it Right:} Angular distribution of the $Z_{\rm c}(3900)^\pm$ polar angle $|\cos\theta_{Z_{\rm c}}|$ in comparison to 
        various $J^P$ assumptions from a simultaneous fit to two data sets at $\sqrt{s}=4.23, 4.26$\,GeV. 
}
       \label{ZcSpinParity_b} 
\end{figure}

Therefore, the spin-parity $J^P$ of $Z_{\rm c}(3885)^\pm\rightarrow D\bar{D^*}^\pm$ has been studied in terms of the angular 
distribution $|\cos\theta_{\pi}|$ between the bachelor pion and the beam axis of the detection efficiency corrected signal 
event yields (Fig.\,\ref{ZcSpinParity_a}). Based on a single $D$-tag analysis (Fig.\,\ref{ZcSpinParity_a},\,left) of data taken 
at 4.26\,GeV, $J^P=1^+$ was determined~\cite{Zc3885_AD_a_BESIII}, and later re-confirmed at higher significance in a 
double $D$-tag analysis based on data at 4.23\,GeV and 4.26\,GeV (Fig.\,\ref{ZcSpinParity_a}, right)~\cite{Zc3885_AD_b_BESIII}. 
Recently, also the spin-parity of the $Z_{\rm c}(3900)^\pm\rightarrow J/\psi\pi^\pm$ system has been studied in an amplitude 
analysis (Fig.\,\ref{ZcSpinParity_b}), including a simultaneous fit to the data sets at 4.23\,GeV and 4.26\,GeV~\cite{Zc3900_PWA_BESIII}. Not only the 
$J^P=1^+$ assignment for this state is significantly favoured by the data, but also the pole mass of 
$(3881.2 \pm 4.2_{\rm stat.} \pm 52.7_{\rm syst.})$\,MeV/$c^2$ is found to be consistent with that of the $Z_{\rm c}(3885)^\pm$ from the 
open charm channel~\cite{Zc3885_AD_b_BESIII}, and this holds also for the ratio 
${\cal B}(Z_{\rm c}^\pm \rightarrow D\bar{D^*}^\pm / \,{\cal B}(Z_{\rm c}^\pm \rightarrow J/\psi\pi^\pm)$ of about $6.2\pm2.9$. 

In conclusion, it seems that these two $Z_{\rm c}$ states at 3.9\,GeV/$c^2$ are indeed the same object observed in different decay modes.
Clearly, also further decay channels via other charmonia than $J/\psi$ and $h_{\rm c}$, like {\it e.g.} $\eta_c$~\cite{FNerling_etal}, 
need to be investigated, and possible multiplets need to be completed also by high-spin states, which can only be accessed by future experiments 
such as PANDA/FAIR~\cite{panda}. 

\subsection{Line-shapes of $J/\psi\pi^+\pi^-$, $h_{\rm c}\pi\pi$, $\psi(2S)\,\pi^0\pi^0/\pi^+\pi^-$ 
and $\pi^+ D^0 D^{*-}$}
The $Y(4260)$ and the $Y(4360)$ were discovered decaying to $J/\psi\pi^+\pi^-$ and $\psi(2S)\pi^+\pi^-$, respectively~\cite{Y4260_barbar,Y4360_barbar}. 
Based on increased statistics, the $Y(4260)$ appears with a somewhat asymmetric shape~\cite{Y4260_barbar_update}. The Belle experiment confirmed the 
$Y(4260) \rightarrow J/\psi\pi^+\pi^-$. In contradiction to the BaBar result, they claimed a lower mass peak, the ``$Y(4008)$'', to be needed in 
addition in order to describe the data~\cite{Y4008_belle}.
\begin{figure}[tp!]
\vspace{-0.3cm}
    \begin{center}
     \includegraphics[clip, trim= 30 140 20 130,width=1.0\linewidth]{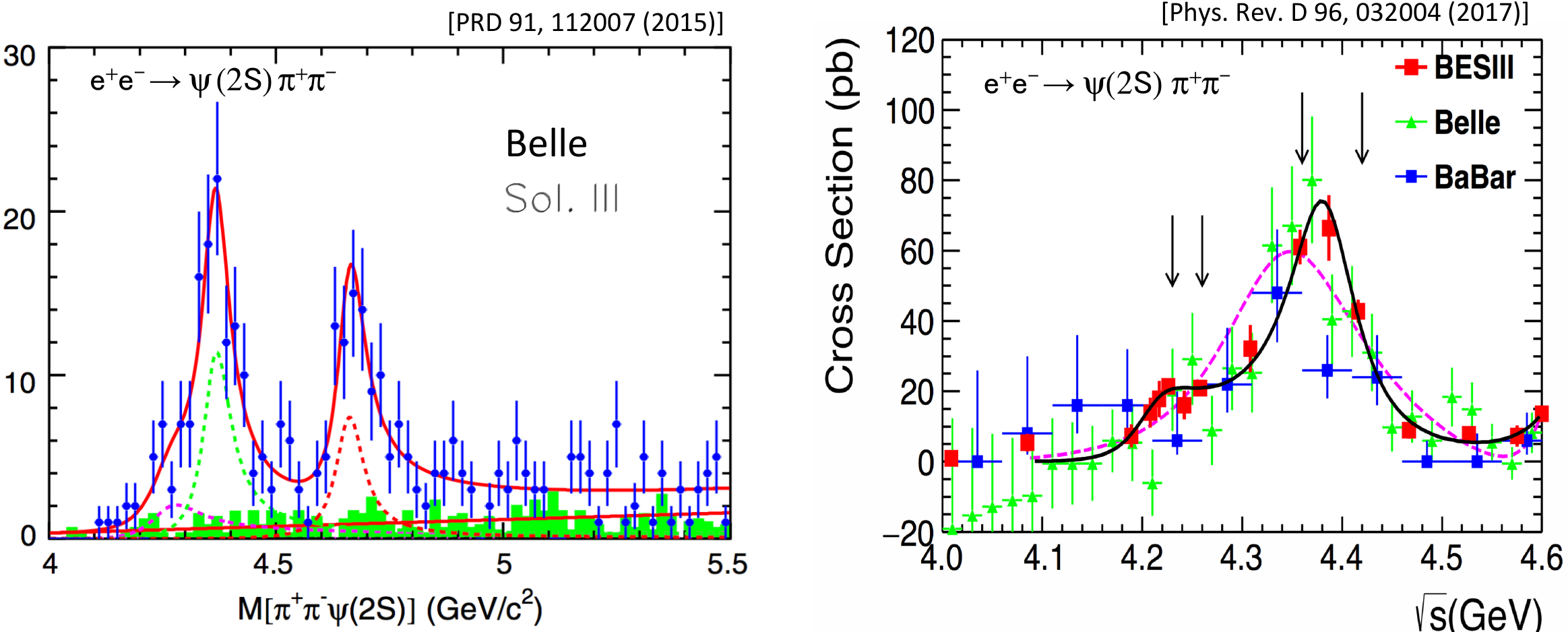}
    \end{center}
    \vspace{-0.6cm}
       \caption{
         {\it Left:} The reconstructed $\psi(2S)\pi^+\pi^-$ invariant mass from Belle shows clear 
         indications of the $Y(4360)$ and the $Y(4660)$, however, no evidence for the $Y(4260)$ resonance.
         {\it Right:} Comparison of the $e^+e^- \rightarrow \psi(2S)\pi^+\pi^-$ cross-section shape 
         as measured at BESIII to those provided by BaBar and Belle -- the $Y(4360)$ line shape is 
         found to be in consistency for the three different experiments. The arrows indicate the four BESIII 
         high luminosity ``XYZ'' data sets.
         }
 \label{Ystates_psi2Spipi_bes3} 
    \vspace{-0.3cm}
\end{figure}
A simultaneous fit to the BESIII high-precision energy dependent cross-section measurements for $\sigma(e^+e^- \rightarrow J/\psi\pi^+\pi^-)$  
of the high luminosity ``XYZ'' (8.2\,fb$^{-1}$) and low luminosity ``$R$-scan'' (0.8\,fb$^{-1}$) data sets~\cite{Ycrosssection_besiii},
resolves two resonance structures at high statistical significance ($>$$7\sigma$) in the $Y(4260)$ region, whereas a $Y(4008)$ appears not to 
be present. 
It should be emphasised that, while the $Y(4260)$ is observed with a significantly smaller width ($\Gamma$=$44.1\pm 3.8$\,MeV/$c^2$) at smaller 
mass ($m$=4220\,MeV/$c^2$), the second resonance (with $m$=$4326.8\pm$10.0\,MeV/$c^2$, $\Gamma$=$98.2^{+25.4}_{-19.6}$\,MeV/$c^2$) 
is (within errors) consistent with the $Y(4360)$, which is here firstly observed in the decay to $J/\psi\pi^+\pi^-$. Previously, it was only 
seen in $\psi(2S)\pi^+\pi^-$~\cite{Y4360_barbar,psi2Spipi_belle}). 

Before coming back to $e^+e^-$ production of the $\psi(2S)\pi^+\pi^-$, the recent BESIII result on $h_{\rm c}\pi^+\pi^-$
production~\cite{hcpipi_besiii} should be mentioned and noted, providing evidence for two resonant structures 
at 4.22\,GeV/$c^2$ ($m$=$(4218.4 \pm4.0 \pm 0.9)$\,MeV/$c^2$ and $\Gamma$=$(66.0 \pm 0.9 \pm 0.4)$\,MeV/$c^2$) and 
at 4.39\,GeV/$c^2$ ($m$=$(4391.6 \pm 6.3 \pm 1.0)$\,MeV/$c^2$, $\Gamma$=$(139.5 \pm 16.1 \pm 0.6)$\,MeV$c^2$) that we 
call ``$Y(4220)$'' and ``$Y(4390)$''. They are observed at a statistical significance of more than 10$\sigma$ over the 
one resonance assumption.

The Belle result on $e^+e^-$ production of $\psi(2S)\pi^+\pi^-$ (Fig.\,\ref{Ystates_psi2Spipi_bes3}, left) shows clear 
indication of $Y(4360)$ and $Y(4660)$, both decaying to $\psi(2S)\pi^+\pi^-$~\cite{psi2Spipi_belle}. However, no 
evidence for the $Y(4260)$ is found in the data ($<$3$\sigma$), and it was thus omitted from their best fit. The new 
$\psi(2S)\pi^+\pi^-$ production cross-section result by BESIII~\cite{psi2Spipi_besiii} is compared to the ones from 
BaBar and Belle in Fig.\,\ref{Ystates_psi2Spipi_bes3} (right). The BESIII measurement confirms the $Y(4360)$ line shape 
reported previously, and from our fit with three coherent Breit-Wigner functions, we observe for the first time 
$Y(4220)\rightarrow\psi(2S)\pi^+\pi^-$ and again $Y(4390)$, which are both consistent in the resonance parameters $(m,\Gamma)$ 
with the two structures that we observe in $h_{\rm c}\pi^+\pi^-$~\cite{hcpipi_besiii}. 
\begin{figure}[tp!]
\vspace{-0.3cm}
    \begin{center}
     \includegraphics[clip, trim= 2 125 5 132,width=1.0\linewidth]{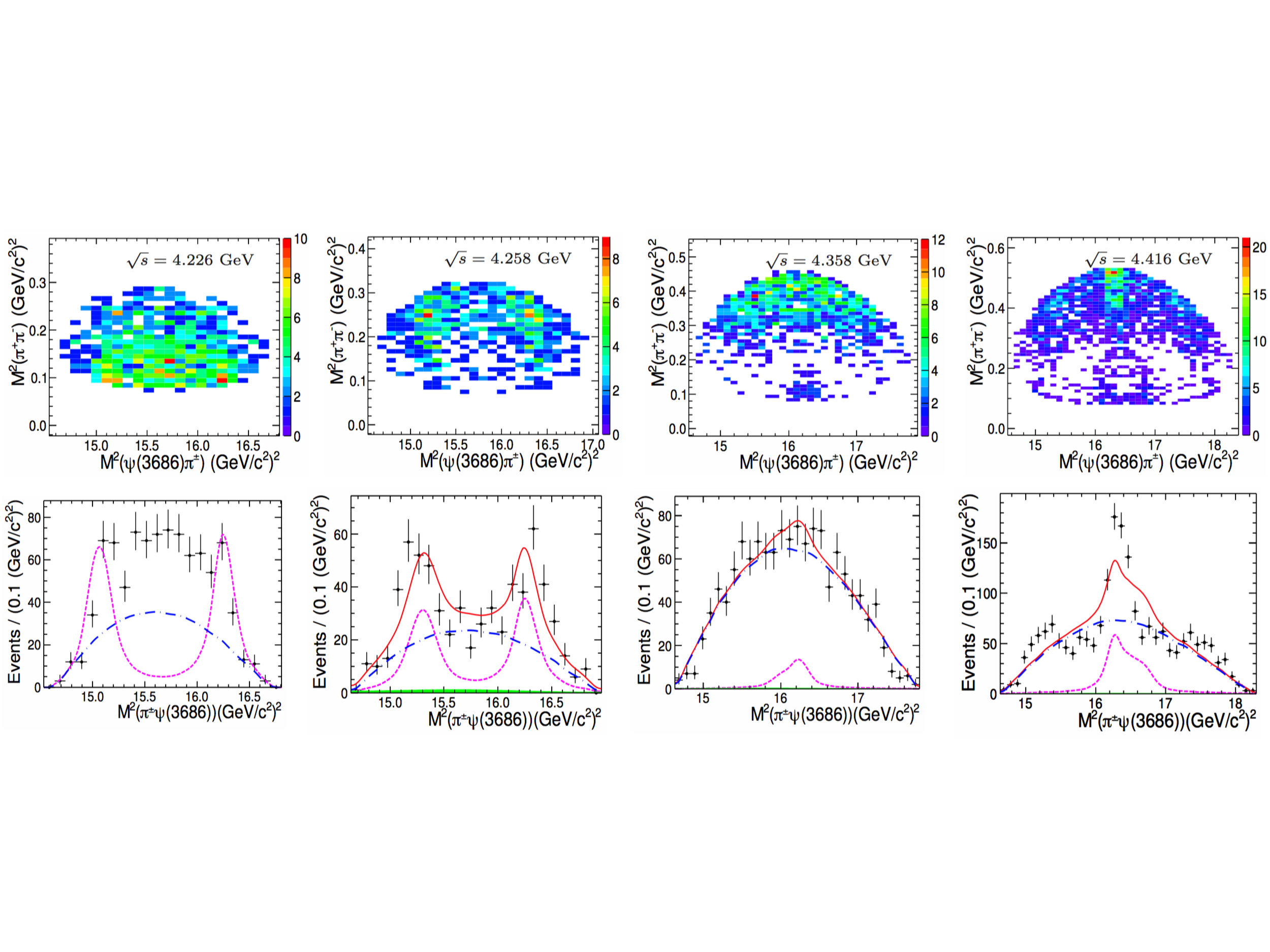}
    \end{center}
\vspace{-0.6cm}
      \caption{{\it Top:} Dalitz plots $m^2(\pi\pi)$ {\it vs.} $m^2(\psi(2S)\pi)$ at the four $E_{\rm cms}$=4.23, 4.26, 4.36 and 4.42\,GeV indicated 
        in Fig.\,\ref{Ystates_psi2Spipi_bes3}.
        {\it Bottom:} Projections of the Dalitz plots to $m^2(\psi(2S)\pi)$ with the fit result overlaid. Both figures taken from~\cite{psi2Spipi_besiii}.  
\vspace{-0.6cm}
}
\label{DPanaBes3psi2Spipi} 
\end{figure}

In the region of these both states, we studied also possible intermediate states using an unbinned maximum-likelihood fit to 
the Dalitz plots (Fig.\,\ref{DPanaBes3psi2Spipi}), in which the parameterisation comprises two coherent sums of resonant and 
non-resonant production. At 4.42\,GeV, including an intermediate state of $(m,\Gamma)$=$(4032.1 \pm 2.4, 26.1 \pm 5.3)$\,MeV/$c^2$ 
improves significantly (9.2$\sigma$) the fit description of the data, consistent with a clearly visible narrow structure in 
$m(\psi(2S)\pi)$ at this $E_{\rm cms}$ (Fig.\,\ref{DPanaBes3psi2Spipi}, top/right). At 4.36\,GeV, no obvious structure is visible 
but a cluster of events at low $m(\pi\pi)$. At the two lower $E_{\rm cms}$ (Fig.\,\ref{DPanaBes3psi2Spipi}, top/left), 
{\it i.e.} in the region of $Y(4220)$, two accumulations of events at about 3.9 and 4.03\,GeV$/c^2$, respectively, are visible 
at $4.26$\,GeV, whereas at $4.23$\,GeV, no structure is clearly seen and the $m(\pi\pi)$ distribution appears very different from 
that at $4.26$\,GeV. It should be noted that possible intermediate states of 3.9 and 4.03\,GeV$/c^2$ at $4.26$\,GeV would have 
kinematic reflections at each other's masses, and at $4.23$\,GeV, a possible $Z_{rm c}(4030)$ state would be rather close to the 
kinematical border, so that no obvious distinct structure would be expected to be visible here. In the fits at 4.36 and 4.26\,GeV/$c^2$, 
the resonance parameters of the intermediate $Z_{\rm c}$-like state were fixed to those obtained at 4.42\,GeV/$c^2$, resulting in 
an statistical significance of 3.6$\sigma$ and 9.6$\sigma$, respectively.  Even though, also at 4.42\,GeV/$c^2$, the data is not 
described sufficiently, the confidence level improves to about 50\,\%, when applying an additional cut of $m(\pi\pi)> 0.3$\,GeV/$c^2$.   

Even though there are still unresolved discrepancies (model {\it vs.} data), we might have observed a $Z_{\rm c}$-like intermediate 
state of a mass of about $m$=$4030$\,MeV/$c^2$. A similar analysis of the neutral counter part, $e^+e^-\rightarrow \psi(2S)\pi^0\pi^0$, 
delivers similar structures and results of the corresponding Dalitz plot analysis~\cite{psi2Spi0pi0_besiii}.
Higher statistics data and theoretical input are needed to improve and sort out the present discrepancies in describing 
the significant sub-structures in the $\psi(2S)\pi\pi$ system. 
\begin{figure}[tp!]
\vspace{-0.7cm}
    \begin{center}
     \includegraphics[clip, trim= 2 25 7 50,width=1.0\linewidth]{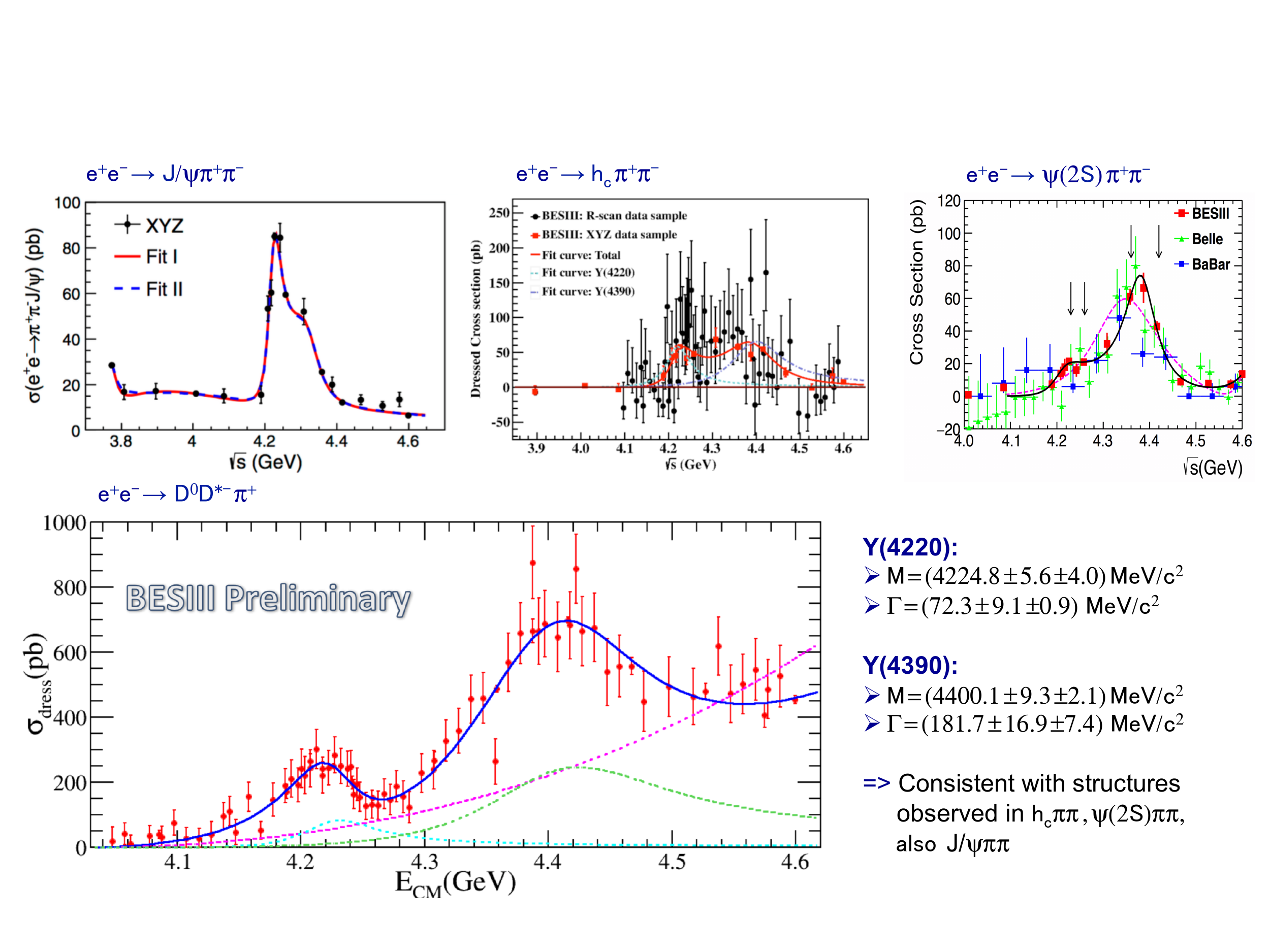}
    \end{center}
\vspace{-0.7cm}
      \caption{Cross-section measurements of direct $J/\psi\pi\pi$, $h_{\rm c}\pi\pi$ and $\psi(2S)\pi\pi$ production, resolving
        consistently the ``$Y(4220)$'' and $Y(4390)$ states ({\it top}) that are also observed in the open charm decay to 
        $D^0D^{*}\pi$ ({\it bottom}). 
}
\vspace{-0.4cm}
       \label{Ystates_Bes3} 
\end{figure}

In conclusion for the vector $Y$ states, we observe two structures, ``$Y(4220)$'' and ``$Y(4390)$'', consistently in the decays 
to $c\bar{c}\pi\pi$, involving the three charmonia $J/\psi$, $h_c$ and $\psi(2S)$, and interestingly, we observe these two 
$Y$ states also being consistent in the resonance parameters as well as in the preliminary open charm analysis of 
$e^+e^-\rightarrow D^0D^{*}\pi$ (Fig.\,\ref{Ystates_Bes3}, bottom). 

\section{Conclusions and outlook}
The BESIII/BEPCII experiment is successfully operating since 2008. Given the world largest data set in the 
$\tau$-charm region, it offers unique possibilities for investigations of the $XYZ$ spectrum. We have the first 
two $Z_{\rm c}$ isospin triplets established, the $X(3872)$ for the first time observed in radiative decays~\cite{X3872_besiii}, 
and we have recently published precision measurements of production cross-section in the $Y$ energy range, resolving for the 
first time structures overseen in previous measurements. Similarly to the $X$ and $Z_{\rm c}$ states, we find these $Y$ states 
also decaying to $DD^*\pi$. As an outlook, BESIII is continuing to collect data, helping and needed to further resolve the 
$XYZ$ puzzle. 
\\\\
{\bf Acknowledgement:} This work is supported by the DFG Grant ``FOR2359''.

\end{document}